\documentclass{article}
\usepackage{amssymb}

\input{tcilatex}

\begin{document}

\title{Kaon-Baryon Couplings and the Goldberger-Treiman Relation in $%
SU(3)\times SU(3)$}
\author{N. F. Nasrallah \\
Faculty of Science, Lebanese University\\
Tripoli, Lebanon\\
nsrallh@ul.edu.lb\\
\linebreak \\
PACS numbers: 11.40Ha,11.55Fv,11.55Hx}
\date{}
\maketitle

\begin{abstract}
The coupling constants $G_{KN\Lambda }$ and $G_{KN\Sigma }$ are obtained from the Goldberger-Treiman relation in the strange channel with chiral symmetry breaking taken into account. The results, $G_{KN\Lambda }=-12.3\pm 1.2$ and $G_{KN\Sigma }=5.5\pm .5$ come close to the $SU(3)$ values.
\end{abstract}

The Kaon-Baryon coupling constants $G_{KN\Lambda }$ and $G_{KN\Sigma }$ are essential ingredients in the study of Kaon-Nucleon scattering or of the strangeness content of the nucleon.

The Goldberger-Treiman relation (GTR) \cite {GT} relates the Meson-Nucleon coupling constants to the axial-vector coupling constant in $\beta -$decay.

In the non-strange channel e.g., the GTR reads
\begin{equation}
m_{N}g_{A}(0)=f_{\pi }G_{\pi N}
\end{equation}%
where $m_{N}$ is the nucleon mass, $g_{A}(0)$ is the axial-vector coupling constant in nucleon $\beta -$decay at vanishing momentum transfer, $f_{\pi }$ is the $\pi $ decay constant and $G_{\pi N}$ is the $\pi -N$ coupling
constant.

The corresponding relations in the strange channel are
\begin{equation}
\sqrt{2}f_{K}G_{KNY}=(m_{N}+m_{Y})g_{A}^{Y}(0)  \label{GKNY}
\end{equation}%
where $Y=\Lambda ,\ \Sigma $.

The $G_{KNY}$ could of course be obtained from eq. (\ref{GKNY}) because the $g_{A}^{Y}(0)$ are measured. Eq. (\ref{GKNY}) however involves an extrapolation in momentum transfer squared from $0$ to $m_{K}^{2}$ which is not as small quantity on the hadronic scale. In other words explicit chiral symmetry breaking by the quark masses lead to corrections to the GTR, the Goldberger-Treiman discrepancies (GTD)\ which are small in the strangeness conserving case \cite{NFN} but which are not so in the strangeness changing one because $m_{s}\gg m_{u,d}$ or alternatively $m_{K}^{2}\gg m_{\pi }^{2}.$

It is therefore desirable to evaluate the corrections introduced by the extrapolation to the GTR in the strange channel.

The GTD's in the strangeness changing case are defined by
\begin{equation}
\Delta ^{Y}=1-\frac{(m_{N}+m_{Y})g_{A}^{Y}(0)}{\sqrt{2}f_{K}G_{KNY}}
\end{equation}

The evaluation of the Kaon-Nucleon coupling constants will yield the GTD's.

Experimentally, the determination of the coupling constants still involves uncertainties \cite{Haberzettl} but they are thought not to differ too much from their $SU(3)$ values
\begin{equation}
G_{KN\Lambda }=\frac{-1}{\sqrt{3}}(3-2\alpha )G_{\pi N}\backsimeq -13.05,\ G_{KN\Sigma }=(2\alpha -1)G_{\pi N}\backsimeq 3.50  \label{GKNL}
\end{equation}
where $\alpha =D/(D+F)=.635$ is the fraction of $D$-type coupling.

The analysis of \cite{Haberzettl} e.g. yields
\begin{equation}
G_{KN\Lambda }=-13.5\ \ \ \ \ \ \ \ \ \ ,\ \ \ \ \ \ \ \ \ \ G_{KN\Sigma}=4.25
\end{equation}

Direct evaluation of $G_{KNY}$ by use of QCD sum rules \cite{BNN} yield values considerably smaller than the ones shown above.

Evaluation of the GTD's using heavy baryon chiral perturbation theory at the one loop level has also been attempted \cite{Goity} resulting in values much smaller than one would expect from eq. (\ref{GKNL}).

It is the purpose of the present work to obtain the $G_{KNY}$ from the GTR with chiral symmetry breaking taken into account.

We start from the matrix element
\begin{equation}
\langle P\left\vert \partial _{\mu }A_{\mu }^{K^{+}}\right\vert Y\rangle=\Pi (q^{2})\overline{P}i\gamma _{5}Y
\end{equation}

$q$ denoting the momentum transfer between the baryons. We have
\begin{equation}
\Pi (0)=(m_{P}+m_{Y)}g_{A}^{Y}(0)  \label{PI(0)}
\end{equation}

Furthermore, the analytic properties of $\Pi (t=q^{2})$ in the complex $t-$plane are known: It has a pole at $t=m_{K}^{2}$ and a cut along the positive $t-$axis starting at $t_{th}=(m_{K}+2m_{\pi })^{2}$:
\begin{equation}
\Pi (t)=\frac{-\sqrt{2}f_{K}m_{K}^{2}G_{KNY}}{(t-m_{K}^{2})}+non-pole\ terms
\end{equation}

Consider now a closed contour $c$ in the complex $t-$plane consisting of a circle of large radius $R$ $(R\thicksim 4-5GeV^{2})$ and two straight lines just above and below the cut and parallel to it extending from threshold to $R$ and the integral
\begin{equation}
\frac{1}{2\pi i}\int_{c}\frac{dt}{t}\Pi (t)=\frac{1}{2\pi i}\int_{th}^{R}\frac{dt}{t}Disc\Pi (t)+\frac{1}{2\pi i}\doint \frac{dt}{t}\Pi (t) \label{Integ1}
\end{equation}

A straightforward application of the residue theorem yields
\begin{equation}
\frac{1}{2\pi i}\int_{c}\frac{dt}{t}\Pi (t)=\Pi (0)-\sqrt{2}f_{K}G_{PKY} \label{Integ2}
\end{equation}

If the integrals appearing in eq. (\ref{Integ1}) were negligible, eqs. (\ref{PI(0)}) and (\ref{Integ2}) would yield the GTR.These integrals are however not expected to be small.

The first integral on the r.h.s. of eq. (\ref{Integ1}) , running along the cut,represents the contribution of the $0^{-}$ strange continuum with quantum numbers of the $K$-meson and provides the main part of the GTD. In the second integral, along the circle, $\Pi ^{QCD}(t)$ can be substituted for $\Pi (t),$an approximation which is expected to be good except possibly for a small region in the vicinity of the positive $t-$axis.

In order to overcome our lack of knowledge of $Disc\Pi (t)$ we consider the modified integral
\begin{equation}
\frac{1}{2\pi i}\int_{c}dt(\frac{1}{t}-a_{0}-a_{1}t)\Pi (t)=\frac{1}{2\pi i}\int_{th}^{R}dt(\frac{1}{t}-a_{0}-a_{1}t)Disc\Pi (t)+\frac{1}{2\pi i}\doint dt(\frac{1}{t}-a_{0}-a_{1}t)\Pi ^{QCD}(t)
\end{equation}

With $a_{0}$ and $a_{1}$ so far arbitrary constants.The residue theorem yields now
\begin{equation}
\frac{1}{2\pi i}\int_{c}dt(\frac{1}{t}-a_{0}-a_{1}t)\Pi (t)=\Pi (0)-\sqrt{2}f_{K}G_{PKY}(1-a_{0}m_{K}^{2}-a_{1}m_{K}^{4})
\end{equation}

The main contribution to the integral over the cut arises from the interval $I:1.5GeV^{2}\lesssim t\lesssim 3.5GeV^{2}$ which includes the resonances (or resonance candidates) $K(1460)$ and $K(1830)$.The constants $a_{0}$ and $a_{1}$are now chosen so as to annihilate the kernel $(\frac{1}{t}-a_{0}-a_{1}t)$ at $t=1.46^{2}GeV^{2}$ and at $t=1.83^{2}GeV^{2}$, i.e.
\begin{equation}
a_{0}=.77GeV^{-2}\ \ \ \ \ \ \ \ \ \ ,\ \ \ \ \ \ \ \ \ \ a_{1}=-.14GeV^{-4}
\end{equation}

With this choice the integrand is reduced to only a few percent of it's
initial value over the interval $I$ and the integral over the cut becomes
negligible. So we have now%
\begin{equation}
\Pi (0)-\sqrt{2}f_{K}G_{KPY}(1-a_{0}m_{K}^{2}-a_{1}m_{K}^{4})\backsimeq \frac{1}{2\pi i}\doint dt(\frac{1}{t}-a_{0}-a_{1}t)\Pi ^{QCD}(t)
\end{equation}

It appears from the equation above that chiral symmetry breaking manifests itself in the presence of the r.h.s.as well as in the deviation of the factor $(1-a_{0}m_{K}^{2}-a_{1}m_{K}^{4})$ from unity . In order to proceed further a knowledge of $\Pi ^{QCD}(t)$ is required.

This knowledge can be obtained from a large number of QCD sum-rule studies \cite{BNN} of the three-point function
\begin{equation}
\Gamma (s=p^{2},t=q^{2})=\int \int dxdye^{-ipx}e^{iqy}\langle 0\left\vert T\Psi ^{P}(x)\partial _{\mu }A_{\mu }(y)\Psi ^{Y}(0)\right\vert 0\rangle
\end{equation}

Where $\Psi ^{P,Y}$ are baryonic currents. We follow the notation of Bracco et al. and set $p^{2}=p^{\prime 2}=s.$
\begin{equation}
\Gamma (s,t)=F(s,t)\sigma _{\mu \nu }\gamma _{5}q_{\mu }p_{\upsilon}^{\prime }+other\ tensor\ structures  \label{Gamma(s,t)}
\end{equation}

Furthermore ,the residue at the double baryonic pole of $\Gamma $ is related to $\Pi (t)$, i.e.

\begin{equation}
\Gamma (s,t)=(\frac{\lambda _{N}\lambda _{Y}\Pi (t)}{(s-m_{N}^{2})(s-m_{Y}^{2})}+...)\sigma _{\mu \nu }\gamma _{5}p_{\mu}^{\prime }q_{\nu }+other\ tensor\ structures  \label{Gamma(s,t)2}
\end{equation}

The $\lambda _{N,Y}$ denoting the couplings of the baryonic currents to the corresponding baryons.

$\Pi ^{QCD}(t)$ is obtained from the above expressions by calculating $F^{QCD}(s,t),$ extrapolating to the baryon mass-shells by use of the Borel transform in the variable $s$ and identifying corresponding terms in eqs. (\ref{Gamma(s,t)}) and (\ref{Gamma(s,t)2}). This gives
\begin{equation}
\Pi _{Y}^{QCD}(t)=\frac{c_{1}^{Y}}{t}+\frac{c_{2}^{Y}}{t^{2}}+...
\end{equation}
with
\begin{equation}
c_{1}^{Y}=\frac{c_{Y}}{\lambda _{N}\lambda _{Y}}\frac{(m_{Y}^{2}-m_{P}^{2})}{(e^{-\frac{m_{N}^{2}}{M^{2}}}-e^{-\frac{m_{Y}^{2}}{M^{2}}})}
\end{equation}
and
\begin{eqnarray}
c_{\Lambda } &\backsimeq &\sqrt{\frac{2}{3}}m_{s}(\frac{4}{3}\langle (\overline{q}q)^{2}\rangle +\frac{4}{3}\langle (\overline{q}q\overline{s}s)\rangle )  \nonumber \\c_{\Sigma } &=&O(m_{s}^{2})\backsimeq 0
\end{eqnarray}

$M^{2}$ is the Borel mass parameter obtained in \cite{BNN}, as well as other threshold parameters , by stability considerations. If we further use the factorization approximation together with the estimates $\langle \overline{s}s\rangle /\langle \overline{q}q\rangle \backsimeq .50$ \ , $\langle m_{s}\overline{s}s\rangle \backsimeq -1.12\ast 10^{-3}MeV^{4}$ \cite{KS} we obtain
\begin{equation}
c_{1}^{\Lambda }\backsimeq .30GeV^{3}\ \ \ \ \ \ \ \ \ \ ,\ \ \ \ \ \ \ \ \ \ c_{1}^{\Sigma }\backsimeq 0
\end{equation}

We then have
\begin{equation}
\Pi ^{Y}(0)-\sqrt{2}f_{K}G_{KNY}(1-a_{0}m_{K}^{2}-a_{1}m_{K}^{4})=a_{0}c_{1}^{Y}+a_{1}c_{2}^{Y}
\end{equation}

$c_{2}^{Y}$ is not available, it will be estimated and taken to represent the error:
\begin{equation}
c_{2}^{Y}\backsim \pm m^{2}c_{1}^{Y}
\end{equation}
with $m^{2}$ a typical hadronic mass, $m^{2}=1GeV^{2}$ say.Thus
\begin{equation}
G_{KNY}=\frac{(m_{N}+m_{Y)}g_{A}^{Y}(0)-a_{0}c_{1}^{Y}-a_{1}c_{2}^{Y}}{\sqrt{2}f_{K}(1-a_{0}m_{K}^{2}-a_{1}m_{K}^{4})}
\end{equation}

Numerically, the error, represented by the last term in the numerator of the equation above, amounts to about 2.5\% of the total. We shall however enlarge it to 10\% in order to account for the uncertainties which arise from the choice of the mass and threshold parameters, factorisation, etc. in the evaluation of $\Pi ^{QCD}(t)$. The values obtained for the coupling constants are thus,with $g_{A}^{\Lambda }(0)=-.72$ and $g_{A}^{\Sigma}(0)=.34$
\begin{equation}
G_{KN\Lambda }=-12.3\pm 1.2\ \ \ \ \ \ \ \ \ \ ,\ \ \ \ \ \ \ \ \ \ G_{KN\Sigma }=5.5\pm .5
\end{equation}

These values do not differ much from the $SU(3)$ \ ones appearing in eq. (\ref{GKNL}). The GTD's are large, as expected

\begin{equation}
\Delta ^{\Lambda }=.25\ \ \ \ \ \ \ \ \ \ ,\ \ \ \ \ \ \ \ \ \ \Delta^{\Sigma }=.18
\end{equation}

In conclusion we have obtained the coupling constants $G_{KNY}$ from the GTR.These come close to their $SU(3)$ vvalues. Chiral symmetry breaking has been taken into account, it is quite large, as shown by the values of the
GTD's.\pagebreak

\end{document}